\documentclass[twocolumn,showpacs,preprintnumbers,amsmath,amssymb]{revtex4}

\usepackage{graphicx}
\usepackage{dcolumn}
\usepackage{bm}

\begin{document}

\preprint{APS/123-QED}

\title{Focusing a 
radially polarized light beam to a significantly smaller spot size}

\author{R. Dorn}%
\author{S. Quabis}
\email{quabis@physik.uni-erlangen.de}

\author{G. Leuchs}
\affiliation{
Max-Planck-Research-Group for Optics, Information and Photonics, Universit\"at Erlangen-N\"urnberg, D-91058 
Erlangen, Germany}

\date{\today}

\begin{abstract}
We experimentally demonstrate for the first time that a radially polarized field can be focussed to a spot size significantly smaller (0.16(1)$\lambda^2$) than for linear polarization (0.26$\lambda^2$). The effect of the vector properties of light is shown by a comparison of the focal intensity distribution for radially and azimuthally polarized input fields. For strong focusing a radially polarized field leads to a longitudinal electric field component at the focus which is sharp and centered at the optical axis. The relative contribution of this component is enhanced by using an annular aperture.
\end{abstract}

\maketitle
A large number of optical instruments and devices make use of a sharply 
focussed light beam.
Prominent examples are in lithography, confocal microscopy, optical data 
storage as well as in particle trapping. A highly concentrated and well 
matched field is also a necessary requirement for manipulating nanoscopic 
quantum systems in quantum information processing or cavity quantum 
electrodynamics \cite{KimbleEnk}. When reaching for the limits the polarization properties 
of the electromagnetic field play a dominant role.
For a vectorial description of strong focusing using high numerical 
apertures different methods have been developed 
\cite{RichardsWolf,MansuripurJOSAA,KantJModOpt,SheppardMultipoleJModOpt}.
It turns out that in general all three mutually orthogonal 
field components appear at the focal region. The electric 
energy density patterns of individual polarization components were 
mapped using single molecules with the absorption dipole axis pointing in 
the three different directions 
\cite{HechtJMicroscopy,NovotnyYoungworthPRL,HechtNovotnyPRL}. For linearly polarized light, 
the energy density distribution of a longitudinally polarized component in 
the direction of propagation of the beam is not rotationally symmetric. 
This primarily causes an asymmetric deformation of the focal spot. When 
using an annular aperture, the relative contribution 
of the longitudinal component is increased and the asymmetry becomes 
more pronounced \cite{SheppardAnnularApert}. These asymmetries were 
recently confirmed experimentally \cite{DornJModOptics}. In addition, it was 
predicted that a special polarization pattern is needed for focusing down 
to the smallest possible spot \cite{QuabisOptComm}.

The effects of the vector properties of light on the structure of the focus 
are best observed if one compares two special input beams with identical 
intensity distribution but different polarization properties. A good
example for two such fields are an azimuthally and a radially polarized 
field with identical doughnut shaped intensity distributions.
When focusing with a high numerical aperture the radially polarized input 
field leads to a strong longitudinal electric field component in the 
vicinity of the focus \cite{ScullyZubairyPRA,QuabisOptComm}. In contrast, 
the azimuthally polarized field generates a strong magnetic field on the 
optical axis \cite{NovotnyQuantumDotSpectroscopy} while the electric field 
is purely transverse and zero at centre \cite{YoungworthOpticsExpress}.
In general, the direction of the polarization may vary largely inside the 
focal spot. 
Especially when sub--wavelength--structures are investigated this has to be 
carefully considered in a quantitative analysis \cite{WilsonOptComm}.
The predicted axial light forces acting on particles in an optical tweezer 
set up are different in a vectorial treatment as compared to theories based 
on paraxial and geometric optics \cite{NussenzveigOptPinzette}.
The specific interaction of the different field components at the focus 
with single molecules can provide information about the orientation of the 
absorption dipole of the molecules 
\cite{HechtJMicroscopy,NovotnyYoungworthPRL}. Although these field 
components may also occur in the vicinity of near field apertures 
\cite{BetzigScience}, a far field technique can be advantageous as it 
avoids a modification of the dynamics of the molecule due to an interaction 
with the probe \cite{XieDunnScience,AmbroseScience}.
For the spectroscopy of magnetic dipole transitions in quantum dots it was 
proposed to use an azimuthally polarized field for the excitation and take 
advantage of the strong magnetic and the vanishing electric field component 
on axis \cite{NovotnyQuantumDotSpectroscopy}.
To couple efficiently to small quantum systems such as single atoms the 
incoming field must be well matched in both its amplitude and phase but also in its 
polarization properties \cite{KimbleEnk}.

The vector properties not only affect the local field direction but also 
the intensity distribution at the focus.
This is experimentally demonstrated here below by comparing highly resolved 
measurements for azimuthally and radially polarized input fields.
The radially polarized beam can have a narrow central peak due to the 
appearance of the strong longitudinal field component that is sharply centered around the optical axis. The experimental results 
reported here confirm for the first time, to the best of our knowledge, that a radially polarized input field leads to the smallest possible spot size in far field focusing observed so far.
Three different types of set up have been described with which a radially or azimuthally polarized doughnut mode can be 
produced \footnote{A radially and azimuthally polarized doughnut mode can 
be described as a superposition of a TEM$_{01}$ and a TEM$_{10}$ Hermite--
Gaussian mode with orthogonal polarization.}. Most of them either generate the mode inside a laser resonator \cite{Pohl,Niziev} or use a Mach--Zehnder like interferometer \cite{Tidwell,YoungworthSPIE}. A third approach involves the use of mode--forming holographic and birefringent elements \cite{Tschudi}.

\begin{figure}[h!]
\includegraphics{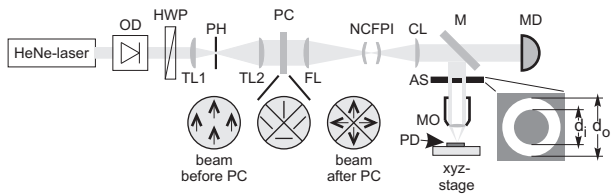}
\caption{\label{fig:Aufbau} Experimental set up to measure the focal 
intensity distribution. MO: Leica plan--apo 100x, 0.9, OD: optical diode, 
HWP: half--wave plate, TL: telescope lenses, PH: pinhole, PC: polarization 
converter, FL: focusing lens, NCFPI: non--confocal Fabry--Perot 
interferometer, CL: collimating lens, AS: aperture stop, M: four mirrors 
for polarization insensitive deflection (details not shown), MD: monitor 
diode, PD: photo diode partially covered with gold--zinc alloy}
\end{figure}

Here we use another approach for the generation of this polarization mode. The TEM$_{00}$--beam of a linearly polarized, single mode 
helium--neon laser ($\lambda$=632.8nm) is expanded and collimated by a 
telescope with an additional pinhole as a mode cleaner (Fig. 
\ref{fig:Aufbau}). The collimated beam is then sent through a polarization 
converter. This element consists of four half--wave plates, one in each 
quadrant. The optical axis of each segment is oriented such that the field 
is rotated to point in the radial direction. However, due to the limited 
number of segments and some diffraction losses at the boundaries between 
adjacent segments, the field is not perfectly radially polarized. 
Nevertheless, this field has a large overlapp with the desired mode and a 
smaller overlapp with additional higher order transverse modes. These 
undesired modes are filtered out by sending the beam through a non--
confocal Fabry--Perot interferometer that is operated as a mode--cleaner 
and which is kept resonant only for the radially or azimuthally polarized 
doughnut--mode. With this method, a purity of $>$99\% is achieved. 

The beam was collimated such that the beam radius at the maximum intensity 
was 1.2mm.
To ensure minimum wave front aberrations we analyzed the beam with a Shack--Hartmann wave front sensor yielding a root mean square wave front error 
$<\lambda/35$. By rotating the polarization of the beam before the polarization converter by 
90$^\circ$, it is possible to switch between a radially and an azimuthally 
polarized beam.
For measurements with an annular aperture, a stop is placed directly in 
front of the microscope objective to block the inner part of the beam. The 
stop is a high quality glass substrate (surface accuracy of $<\lambda/20$) 
coated with an opaque disc (d$_{i}$=3.0mm or 3.3mm). The diameter of the entrance 
pupil of the focusing microscope objective (NA=0.9) is 3.6mm.
To measure the focal intensity distribution \footnote{In this paper, 
intensity always refers to electric energy density, which is the part of 
the field energy that couples to standard photodetectors and photosensitive 
materials.} we used an experimental set up which is based on the knife edge 
method. The edge is formed by a sharp edged opaque pad that is deposited on 
the active surface of a photodiode. Therefore no subsequent optical system 
is needed to collect the transmitted light and effects due to diffraction 
at the edge are minimized. A more detailed description of the properties of 
the sensor is given in \cite{DornJModOptics}.
The knife edge method allows one to measure the one dimensional 
projection i.e. the line integrated intensity of the two dimensional 
intensity distribution onto the direction along which the edge is moved. By 
measuring a set of projections onto different directions one acquires the 
input data needed for a tomographic reconstruction of the two dimensional 
intensity distribution, using e.g. the Radon back--transformation formalism 
\cite{ImageReconstruction}.

\begin{figure}[h!]
\includegraphics{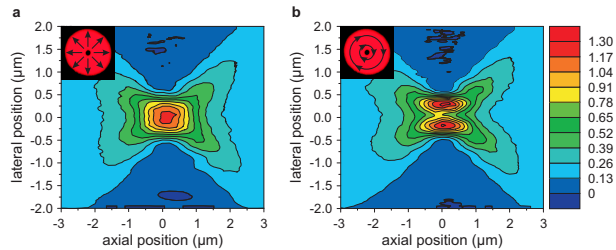}
\caption{\label{fig:ZSerienRadAzi_2} Defocus series for a radially 
(a) and azimuthally (b) polarized input field. The insets illustrate how 
the polarization of the input beam varies spatially across the beam cross 
section. Note that identical initial intensity distributions lead to 
different distributions at focus.}
\end{figure}

For rotationally symmetric input fields the focal spot also shows 
rotational symmetry and all projections onto any line are identical.
Therefore one representative measurement would in principle suffice. 
But we measured projections onto two orthogonal lines to exclude the 
possibility that the symmetry of the focussed field is affected by non--spherical aberrations introduced e.g. by the microscope objective. The 
results were identical within the limits of reproducibility.
Line integrals of the intensity distribution were also measured for 
different positions of the knife edge sensor with respect to the focal 
plane. This yields a line integrated cross section of the propagating beam 
containing the optical axis (Fig. \ref{fig:ZSerienRadAzi_2}).

\begin{figure}[h!]
\includegraphics{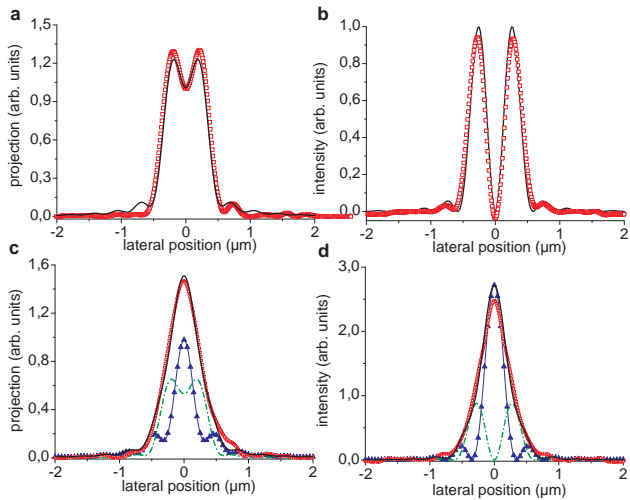}
\caption{\label{fig:ProjURekOhneRing}Measured projections (a,c) and cross sections through
the tomographically reconstructed intensity distributions (b,d) in the focal 
plane for an azimuthally (a,b) and a radially (c,d) polarized input field. Open squares: experimental data; calculated curves: solid black line: total field; dashed line: transverse field; solid triangles: longitudinal field}
\end{figure}

An evidence for the strong longitudinal electric field at the focus of the radially polarized beam is the fact that the intensity projection shows a maximum on the optical axis in the focal plane (Fig. \ref{fig:ZSerienRadAzi_2}a). In contrast, the azimuthally polarized beam has no longitudinal field component and on the optical axis the intensity projection has a minimum. The measured projections in the focal plane are shown in figure \ref{fig:ProjURekOhneRing} for the two modes along with the calculated distributions of the transverse and longitudinal field components. The curves were normalized such that for each case the total intensity in the focal spot equals 1 (arbitrary unit).
\noindent Cross sections for the tomographically reconstructed intensity 
distributions (Fig. \ref{fig:ProjURekOhneRing}) show that 
the intensity on the optical axis vanishes for the azimuthally polarized 
beam.
The beam profile at the focus is very similar to the doughnut 
shaped profile of the input beam just as one would naively expect based on 
the scalar diffraction theory.
\\
In contrast, the radially 
polarized mode leads to a narrow peak around the optical axis where the 
field is basically directed into a longitudinal direction. The transverse 
field components show a profile very similar to the profile of the 
azimuthally polarized beam. The power contained in the longitudinal field 
is calculated to be 49.6\% of the total beam power. The relative 
contribution of the longitudinal component can be enhanced when the 
numerical aperture of the focusing system is increased.\\
For a numerical aperture of 0.9 as in this experiment the calculated spot size \footnote{The spot size is defined as the area that is encircled by the contour line at half the maximum value of the intensity.} associated with this longitudinal field alone is 0.2$\lambda^2$ which is below the spot size for a linearly polarized input field with homogeneous intensity distribution (0.31$\lambda^2$). The spot size of the total field is still larger (0.47$\lambda^2$) than for linear polarization due to the broad distribution of the additional transverse field component. We note in passing that, in general, a calculation using a scalar field tends to underestimate the spot size for large numerical apertures. The reason is that  due to the vector properties of light 
the plane waves that form the focussed field in image space cannot 
interfere perfectly at the focus as the electric (and magnetic) field 
vectors are not all parallel at the focus.

In the case of the radially polarized doughnut mode the rays that propagate 
under a small angle to the optical axis show only a small longitudinal 
component. The stronger transverse field components associated with these 
waves all cancel out on the optical axis and form a broad, doughnut shaped 
profile which increases the spot size. But when an 
annular aperture is used for focusing, only waves that propagate under a 
large angle to the optical axis contribute to the focal field. For a high numerical aperture and a large inner radius  of the annular aperture the electric field vectors for a radially polarized 
\begin{figure}[h!]
\includegraphics{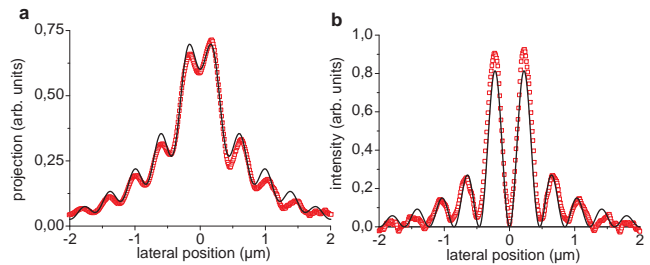}
\caption{\label{fig:AziMitRingb} Projection of (a) and cross section through (b) the tomographically reconstructed intensity distribution for an azimuthally polarized input field focussed using an annular 
aperture (d$_{i}$=3.0mm). Open squares: experimental data; solid black line: calculated curve for the 
total field}
\end{figure}
\begin{figure}[h!]
\includegraphics{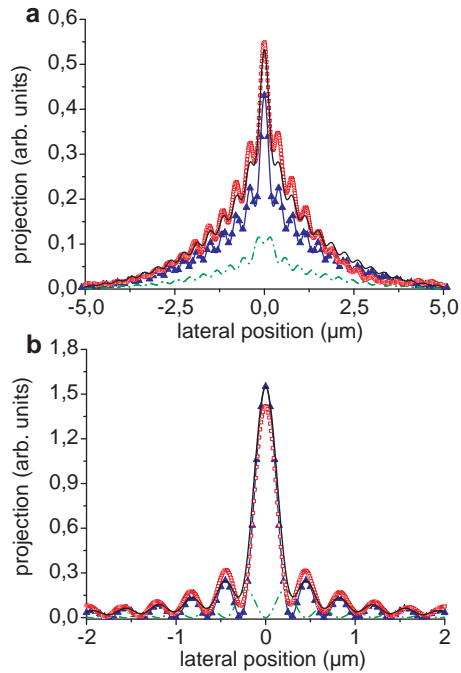}
\caption{\label{fig:RadMitRingb} Measured projection of (a) and cross section through (b) the tomographically reconstructed intensity distribution for a radially polarized input field focussed using an annular aperture (d$_{i}$=3.3mm). The calculated power contained in the longitudinal 
field amounts to 72.8\% of the total beam power.
Open squares: experimental data; calculated curves: solid black line: 
total field; dashed line: transverse field; solid 
triangles: longitudinal field}
\end{figure} 
 input field are essentially 
parallel to the optical axis. All rays interfere perfectly and consequently 
the spot size is close to the spot size which one would calculate using a 
scalar theory. 
Cross sections through the tomographically reconstructed intensity distributions are shown for an azimuthally (Fig. \ref{fig:AziMitRingb}) and for a radially (Fig. \ref{fig:RadMitRingb}) polarized input field when using the annular aperture. Compared to the case 
without annular aperture the sidelobes are more pronounced in both cases, 
indicating that the relative contribution of the intensity in the side 
lobes has increased at the expense of a decreased central maximum.

The spot size for radial polarization with annular aperture is reduced to 0.16(1)$\lambda^2$ 
(theoretical value 0.17$\lambda^2$) (Fig. \ref{fig:RadMitRingb}b). This is well below 
0.26$\lambda^2$, the theoretically achievable spot size for linear 
polarization under the same experimental conditions and also well below 
0.22$\lambda^2$, the theoretical value for circularly polarized light. 
The latter is a little smaller as compared to linear polarization due to the definition of the spot size used here.

Expressed in units of $\lambda^2$ this is to our knowledge the smallest 
measured spot size that has been reached in far field focusing in air 
(NA$<$1).

It is worth noting that for a radially polarized input field, the magnetic 
field vectors point in the azimuthal direction just as the electric field 
vectors do for an azimuthally polarized input beam. The magnetic energy 
density distribution for a radially polarized beam shows the same 
distribution as the electric energy density distribution for an azimuthally 
polarized beam and vice versa.

In summary, we have experimentally verified the influence of the 
polarization on the shape of the focal spot. For a radially polarized field 
distribution with annular aperture, the focal spot size was reduced to the 
smallest focal spot observed up to now when normalizing to the wavelength 
($\lambda^2$). The experimentally observed spot size for NA=0.9 is about 
35\% below the theoretical limit for linearly polarized light.
The calculated spot size for the longitudinal field alone is 
0.14$\lambda^2$. Therefore, if a surface is covered with a special 
photosensitive layer which is only sensitive to the narrow longitudinal 
field distribution \cite{QuabisOptComm} even smaller spot sizes can be 
reached.
These ideas may well be combined with other concepts for the increase of 
transverse resolution such as the solid immersion lens \cite{KinoSIL} or coupling to small mesoscopic antennae \cite{BouhelierEtAlPRL}. A 
radially polarized field will also provide the input field which is best 
suited for coupling into the recently proposed all--dielectric waveguide 
\cite{IbanescuScience}.
\begin{acknowledgments}
We thank G. D\"ohler, S. Malzer and M. Schardt for the fabrication of the p-i-n diode.
We appreciate helpful discussions with H.\,J.\,Kimble and 
C.\,J.\,R.\,Sheppard. We thank J.\,Pfund for the wavefront measurements with
the Shack-Hartmann sensor. This work was supported by the EU grant under QIPC, Project No. IST-1999-13071 (QUICOV).
\end{acknowledgments}

\end{document}